\documentstyle[amssymb,prb,twocolumn,epsf,aps]{revtex}

\begin{document}
\draft

\twocolumn[\hsize\textwidth\columnwidth\hsize\csname
@twocolumnfalse\endcsname

\title{Partially confined excitons in semiconductor nanocrystals with a finite size
dielectric interface}
\author{P. G. Bolcatto\cite{dir} and C. R. Proetto}
\address{Comisi\'{o}n Nacional de Energ\'{\i}a At\'{o}mica\\
Centro At\'{o}mico Bariloche and Instituto Balseiro\\
8400 Bariloche, Argentina}
\maketitle

\begin{abstract}
The combined effect of finite potential barriers and dielectric mismatch
between dot and matrix on excitonic properties of semiconductor quantum dots
has been studied. To avoid the unphysical divergence in the
self-polarization energy which arises for the simplest and profusely adopted
step-like model of the dielectric interface, we proposed a realistic (finite
size) smooth profile for the dielectric interface. We have found that the
excitonic binding energy can be either higher than the corresponding one to
complete confinement by infinite barriers or essentially zero for a wide
range of dot sizes depending on the thickness of the dielectric interface.
\end{abstract}


\vskip2pc]

\narrowtext 

\section{Introduction}

The increasing experimental capabilities in the fabrication of semiconductor
nano-crystallites (quantum dots, QD) for a wide spectra of sizes drives the
experimental and theoretical interest for the overall comprehension of these
systems since, due to the quantum confinement, the electronic and optical
properties are strongly dot size-dependent.\cite{yoffe} In particular, it is
well established that the optical band-gap energy is augmented (as compared
with bulk values) towards higher energies when the dot size decreases. This
behavior is the result of two opposite effects: while the single-particle
band gap energy is shifted to higher energies, the binding energy of the
electron-hole pair (exciton) created by photoexcitation adds an attractive
correction which gives rise to a net blue shift but with a weaker dot size
dependence as compared with the corresponding to the single-particle band
gap energy.

The first and simplest theoretical approach used in the analysis of
electronic properties of nanostructures was the effective mass approximation
(EMA),\cite{efros,brus} which in principle is capable of covering the full
range from large to small quantum dots, with the length scale given roughly
by a few lattice constants. It is then clear that there should exist a
critical dot size for the validity of the EMA, below which the approximation
(essentially based in the semiconductor bulk properties) must fail.
Fortunately, in this limit of very small crystallites (up to a maximum
diameter of about 30 \AA ) other more microscopic approaches such as the
empirical tight-binding (ETB),\cite{lannoo,proot,lannoo1,hill} the empirical
pseudopotential method (EPM),\cite{rama,wang,zunger} and {\it ab initio }%
pseudo-potential\cite{luie} calculations are possible and improvements in
quantitative agreement with experimental results has been reached in this
small size regime. Based in a comparison between their results for the
single-particle band gap energy, the Coulomb and exchange excitonic
corrections and the corresponding EMA results, most of the authors of this
group conclude in disregard the EMA as a quantitative, and even qualitative
method for the study of nanocrystallites. However, we must put attention to
the fact that in most of the cases the EMA results were obtained under the
assumption of perfect confinement (infinite barriers, IEMA) for both
electron and hole. Several authors pointed out that this extreme hard-wall
constraint (which is not inherent to this approach) is the main reason for
the discrepancy with more microscopic approaches and carried out
calculations assuming an incomplete confinement (finite barriers, FEMA) for
the electron and hole.\cite{kayanuma,fema} The output of such calculations
is that while the single-particle band gap energy is not well described by
FEMA, excellent quantitative agreement was obtained for the excitonic
properties with the more sophisticated calculations available at present .
This can be understood from the fact that while single-particle properties
depend directly on the fine structure of the electron and hole envelope
functions, exciton properties are given by averaged (integrated) expectation
values with these envelope functions (see below). As a consequence,
excitonic related properties are much less sensitive to fine details of the
electron and hole wave functions than single particle properties. The
encouraging results obtained in this way allow us to keep FEMA as a
qualitative, flexible and versatile theoretical tool for the study of the
excitonic physics of quantum dots.

Another question to analyze in detail in order to be able to make a proper
comparison between theory and experiment is the change in the excitonic
energy due to the dielectric mismatch between the quantum dot and the
surrounding matrix. This brings immediately the issue that electrostatic
screening in semiconductor quantum dots is expected to be different from its
bulk counterpart. In particular, reasonable doubts arises about if a
macroscopic treatment based on the use of an effective dot dielectric
constant $(\varepsilon )$ is still meaningful. This topic was recently
addressed in three different calculations, a phenomenological one,\cite{tsu}
and two full quantum mechanical ones.\cite{lannoo1,wang1} All of them
conclude that the confinement should reduce $\varepsilon $, although
differed with respect to the magnitude of this reduction. Besides, and this
is the key point, in the last two microscopic calculations, through a
careful comparison between microscopic and macroscopic results for the
electrostatic screening, the validity of macroscopic type treatments was
established, although with suitable generalizations. This size-dependent
effect on the dot dielectric response can be easily incorporated to our
approach, but we shown at the end of Section III that this correction is
quantitatively quite small as compared with the size-dependence effects on
the Coulomb and self-energies. Consequently, in the rest of the paper we
ascribe to the dot a bulk dielectric constant, which is an excellent
approximation for not too small quantum dots $($dot diameter, $d\gtrsim 20$ 
\AA $)$. For simplicity, most of the times the effect of the dielectric
mismatch is incorporated assuming that the dielectric constant changes
discontinuously from its value in the quantum dot to the value corresponding
to the matrix (step-like model, see Fig. 1a). In this case, and according 
to the classical
image-charge concepts of basic electrostatics,\cite{jackson} the presence of
a charged particle inside the quantum dot induces charges at the boundaries
 which have the same sign as the source charge if the
dielectric constant inside the dot is higher than outside (the general
situation). Within IEMA, this gives rise to a repulsive (positive) potential
energy corresponding to the particle interacting with its own induced
charge; this is usually denoted as a self-polarization energy correction. 
Besides, when we study excitons, the polarization effects also modify the 
direct Coulomb interaction since the induced charge by the electron (hole) 
can attractively interacts with the hole (electron) for the usual case. 
Consequently, there are two opposite contributions to the excitonic energy
leaded by polarization effects. For spherical quantum dots, and assuming i) 
infinite confinement for the particles (IEMA) and, ii) a step-like model for 
the dielectric profile, it is possible to demonstrate\cite{lanoo1} that both 
effects are almost exactly compensated and then, the exitonic energy is reduced
to the bare electron-hole attraction. In a previous work, we have verified that 
the same cancelation occurs for cubic quantum dots.\cite{shape} However, as soon 
as we depart from this two strong assumptions, the compensation between the
polarization effects will not verify anymore. Therefore, the two polarization 
contributions to the energy must be taked into account in more realistic
model calculations.

A direct consequence of the step-like dielectric profile is that all the induced 
charge is perfectly localized at the
surface boundary, leading to a divergent self-polarization potential at the
dot surface. This feature is not compatible with the electron and hole
incomplete confinement associated with finite potential barriers, since in
this case both the particle and its induced charge could coincide at the dot
boundary and then, the self-polarization energy diverges.\cite{limite} This
unphysical result clearly show that the standard electrostatic description
of the interface between dielectric media fails at distances comparable to
the interatomic distance. The difficulty is automatically solved in the
infinite barrier case: as a consequence of the hard-wall boundary condition
the electron and hole envelope wave functions are zero precisely where the
self-induced potential diverges (the dot boundary), and then the
self-polarization energy remains finite. For the finite barrier case, the
anomalous behavior can be ``regularized'' by introducing a phenomenological
cut-off distance of the same order of magnitude as the interatomic distance.%
\cite{kumagai,tran,ban} However, this proposal of modifying the functional
form of the potential does not follow from any calculation; it is only a
mathematical trade-off designed to avoid the divergence. A more rigorous
approach was carried out by Stern\cite{stern} who, in the study of a
semiconductor-insulator Si/SiO$_{2}$ planar interface, solves the problem
proposing a model dielectric interface in which the dielectric constant
changes continuously between its extreme values (see Fig. 1b for our
three-dimensional analogue). In this way the induced charge is spread along
the dielectric interface region and the divergence in the self-polarization
energy disappears, as it physically should be. The consideration of an
interface with finite thickness is in parallel with the situation in real
samples because the experimental growth procedures do not guarantee a sharp
profile without interdiffusion between dot and matrix materials.

\begin{figure}[tbp]
\centering
\epsfxsize=6.75cm
\leavevmode
\epsffile{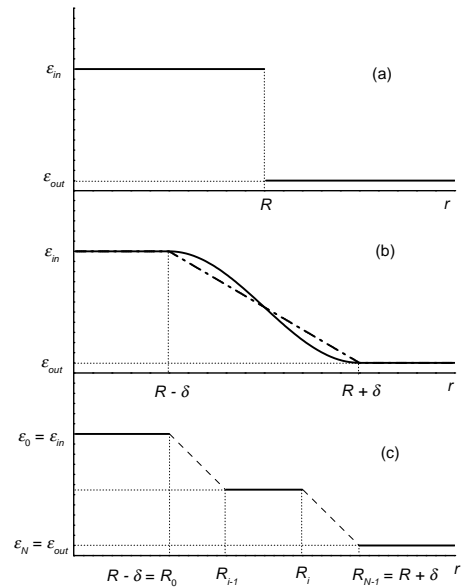}
\vskip1mm
\caption{Spatial profiles for the dielectric interface. (a) Step-like model;
(b) linear (dot-dashed line) and cosine-like (full line) models; (c)
schematic discretization by $N-$ steps of an arbitrary model of the
dielectric interface. The step-like model corresponds to the choice $\delta
=0,$ $N=1$, while in the limit $N\rightarrow \infty $ the profile of Fig. 1c
can be made arbitrarely close to any of the continous profiles of Fig. 1b.}
\label{fig1}
\end{figure}

As is discussed above, the evaluation of the optical band gap requires two
independent calculations: the single-particle band gap energy (blue shift)
and the excitonic corrections (red shift). By using EMA, the accurate
calculation of the former contribution requires consideration of other
ingredients as a multiband ${\bf k}\bullet {\bf p}${\bf \ }finite barrier
calculation of the dot band structure,\cite{bastard1} which is beyond the
scope of the present work. Therefore, the aim of the present paper is to
give insights about the excitonic correction in quantum dots when the
combined effects of partial confinement by finite barriers and dielectric
mismatch are considered.

The rest of the paper is organized as follows: in Section II we give the
basic formulae used for the calculation of the excitonic energies, in
Section III we discuss the results, and finally in Section IV we summarize
our conclusions.

\section{Theory}

In a previous work\cite{shape} we have demonstrated that the excitonic
properties (even in the presence of large dielectric mismatch) are
essentially independent of the dot shape, so we will concentrate our
analysis in the simplest geometry: spherical quantum dots with radius $R$.
We attach to the dot-acting semiconductor a dielectric constant $\varepsilon
_{in},$ while the outside media where the dot is embedded has a dielectric
constant $\varepsilon _{out},$ and we will study the effect of different
models for the dielectric interface connecting both limits (see Fig. 1).
Within the envelope wave function approach to the effective mass
approximation,\cite{bastard} the relevant Hamiltonian for an electron-hole
system is given by 
\begin{equation}
H\left( {\bf r}_{e},{\bf r}_{h}\right) =H_{e}\left( {\bf r}_{e}\right)
+H_{h}\left( {\bf r}_{h}\right) +V_{c}\left( {\bf r}_{e},{\bf r}_{h}\right) ,
\label{echam}
\end{equation}
where the first and second terms on the right-hand side are the
single-particle contributions for the electron and hole respectively, and
the last one is the generalized electron-hole Coulomb interaction (to be
discussed below). The single-particle Hamiltonian $H_{i}\left( {\bf r}%
_{i}\right) $ $\left( i=e,h\right) $ is defined as 
\begin{equation}
H_{i}\left( {\bf r}_{i}\right) =-\frac{\hbar ^{2}}{2m_{i}\left( {\bf r}%
_{i}\right) }\nabla _{i}^{2}+V_{s}\left( {\bf r}_{i}\right) +V_{0i}\left( 
{\bf r}_{i}\right) ,  \label{eckine}
\end{equation}
where the first term is the kinetic energy for a particle with effective
mass $m_{i}\left( {\bf r}_{i}\right) $, $V_{s}\left( {\bf r}_{i}\right) $
corresponds to the above discussed self-polarization energy that can be
obtained as a particular case of $V_{c}({\bf r}_{e},{\bf r}_{h})$ (see
below), and the last one is the confining potential which we assume as a
spherical quantum well-like potential defined by $V_{0i}\left( {\bf r}%
_{i}\right) \equiv \Theta \left( r_{i}-R\right) V_{0i}{\bf ,}$ being $\Theta 
$ the step function and $V_{0i}$ the barrier height. The dependence of $%
m_{i} $ on $r_{i}$ arises from the fact that the particles have different
effective masses depending on their location, inside or outside the dot.

\subsection{Coulomb and self-polarization potentials}

The generalized Coulomb potential between two particles with coordinates $%
{\bf r}$ and ${\bf r}^{\prime },$ and charges $e$ and $e^{\prime }$
respectively, can be calculated through $V_{c}\left( {\bf r},{\bf r}^{\prime
}\right) =e$ $\Phi \left( {\bf r},{\bf r}^{\prime }\right) $, where $\Phi
\left( {\bf r},{\bf r}^{\prime }\right) $ is the electrostatic potential
which verifies the Poisson equation 
\begin{equation}
\nabla _{{\bf r}}\cdot \varepsilon \left( {\bf r}\right) \nabla _{{\bf r}%
}\Phi \left( {\bf r,r}^{\prime }\right) =-4\pi e^{\prime }\delta \left( {\bf %
r-r}^{\prime }\right) ,  \label{ecpoison}
\end{equation}
where $\varepsilon ({\bf r})$ is the (local) spatially dependent dielectric
function. This equation is quite simple to solve analytically if we assume
that $\varepsilon ({\bf r})$ depends only on $r,$ and also that the spatial
dependence for $\varepsilon (r)$ corresponds to the step-like model, as
shown in Fig. 1a (for instance, see Refs. 1b and 3). The important points in
the derivation of such solution are: first, under the assumption of
spherical symmetry Eq.(3) is separable in angular and radial parts, being
only the resulting radial differential equation the non-trivial part to
solve; second, the solution of this radial equation is quite easy to find if 
$\varepsilon (r)$ is constant. For this particular case, the radial solution
of Eq. (\ref{ecpoison}) consists of product of powers of the coordinates $r$
and{\bf \ }$r^{\prime }$; the step-like model of Fig. 1a allows then to
propose for each region at both sides of the interface two different
solutions of this type, and finally obtain the solution for the whole space
through the standard electrostatic boundary conditions which $\Phi $ should
satisfy at the interface. It is also possible to obtain an analytical
solution of Eq. (\ref{ecpoison}) in the case where $\varepsilon \left( {\bf r%
}\right) $ depends linearly on only one coordinate, we say $z$, and is
independent of $x$ and $y$, as for example in the case of a simple
semiconductor-insulator junction.\cite{stern} The disappointing feature of
this exact solution is that the electrostatic potential exhibits a
logarithmic divergence at the edges of the transition layer, which can be
traced back to the discontinuous first derivative at these points of the
linear model for the dielectric interface. The three-dimensional solution
for the linear model of the dielectric interface, even assuming spherical
symmetry, is quite complicated to find since the Eq. (\ref{ecpoison})
becomes a second-order differential equation with spatially dependent
coefficients. We will not pursue the analytical resolution of this problem,
which even if feasible, is restricted to just a particular model for the
dielectric interface.

Instead, by exploiting the exact solution of the single-step case, and
following the spirit of Ref. [\onlinecite{stern}], we will consider an {\it %
arbitrary} smooth radial profile for $\varepsilon \left( r\right) $ and
proceed as follows: first we define a finite interface of width $2\delta $
centered at $R$. For $r<R-\delta $ (well inside the dot) the dielectric
constant takes the value corresponding to the bulk dot-acting semiconductor, 
$\varepsilon \left( r\right) \equiv \varepsilon _{in}.$ For $r>R+\delta $
(well outside the dot) $\varepsilon \left( r\right) $ is fixed at the value
corresponding to the surrounding matrix, $\varepsilon \left( r\right) \equiv
\varepsilon _{out}$. Between them, for $R-\delta $ $<r<R+\delta ,$ we can
choose any analytical and physically plausible function that connects these
extreme values as is pictured in Fig. 1b. Afterwards we subdivide the
dielectric interface in $N-1$ regions and in each one of them we approximate
the dielectric function by a constant value (typically the mean value of the
selected function in this zone). Thus, the original continuous profile is
approximated by an $N$-step one in which each single-step problem has a
known solution. This approach allows us to connect $\varepsilon _{in}$ and $%
\varepsilon _{out}$ with any arbitrary function being the number of steps $N$
the only parameter to adjust, the condition being that it should provide a
good approximation to the true connecting function. Strictly, our approach
is such that when $N\rightarrow \infty $ we recover the exact solution of
Eq. (3). As a result of this procedure, the expression for the generalized
Coulomb potential is (see Fig. 1c to clarify the notation)

\begin{equation}
V_{c}^{i,m}\left( {\bf r},{\bf r}^{\prime }\right) =\frac{ee^{\prime }}{%
\varepsilon _{m}}\text{ }\sum\limits_{l=0}^{\infty }P_{l}\left( \cos \gamma
\right) \frac{F_{im,l}\left( {\bf r},{\bf r}^{\prime }\right) }{\left(
1-p_{m,l}q_{m,l}\right) },  \label{ecvc}
\end{equation}
where the superindex $i$ ( $m$ ) indicates that the particle with coordinate 
${\bf r}\ $( ${\bf r}^{\prime }$ ) is placed at the $i-$region ( $m-$%
region); $P_{l}\left( \cos \gamma \right) $ are the Legendre polynomials of
order $l$ and $\gamma $ the angle between both particles (measured from the
origin at the dot center). With $i$ and $m$ taking values between $0$ and $%
N, $ the Coulomb potential $V_{c}({\bf r},{\bf r}^{\prime })$ is defined in
the whole space. The functions $F_{im,l}\left( {\bf r},{\bf r}^{\prime
}\right) $ are given by 
\begin{eqnarray}
F_{im,l}\left( {\bf r},{\bf r}^{\prime }\right) &=&\left[
p_{m,l}r_{>}^{l}+r_{>}^{-\left( l+1\right) }\right] \left[
r_{<}^{l}+q_{i,l}r_{<}^{-\left( l+1\right) }\right] \times  \nonumber \\
&&\prod\limits_{j=i}^{m-1}\frac{1+q_{j+1,l}R_{j}^{-\left( 2l+1\right) }}{%
1+q_{j,l}R_{j}^{-\left( 2l+1\right) }},
\end{eqnarray}
if $i<m,$%
\begin{equation}
F_{im,l}\left( {\bf r},{\bf r}^{\prime }\right) =\left[
p_{m,l}r_{>}^{l}+r_{>}^{-\left( l+1\right) }\right] \left[
r_{<}^{l}+q_{m,l}r_{<}^{-\left( l+1\right) }\right] ,  \label{ecfmm}
\end{equation}
for $i=m$ and 
\begin{eqnarray}
F_{im,l}\left( {\bf r},{\bf r}^{\prime }\right) &=&\left[
p_{i,l}r_{>}^{l}+r_{>}^{-\left( l+1\right) }\right] \left[
r_{<}^{l}+q_{m,l}r_{<}^{-\left( l+1\right) }\right] \times  \nonumber \\
&&\prod\limits_{j=m+1}^{i}\frac{1+p_{j-1,l}R_{j-1}^{2l+1}}{%
1+p_{j,l}R_{j-1}^{2l+1}},
\end{eqnarray}
when $i>m.$ In the previous equations $r_{<}\left( r_{>}\right) $ is the
smallest (greatest) distance between $r$ and $r^{\prime }.$ The $q_{i,l}$
coefficients are defined so that $q_{0,l}=0\left( \forall l\right) $ and for 
$0<i\leq m$ are expressed through the following recursive expressions 
\begin{equation}
q_{i,l}=\frac{S_{21}^{i,l}+q_{i-1,l}S_{22}^{i,l}}{%
S_{11}^{i,l}+q_{i-1,l}S_{12}^{i,l}},
\end{equation}
where the ${\bf S}-$matrix is 
\begin{equation}
{\bf S}^{i,l}{\bf =}\left[ 
\begin{array}{cc}
\varepsilon _{i}\left( l+1\right) +\varepsilon _{i-1}l & \left( l+1\right)
\left( \varepsilon _{i}-\varepsilon _{i-1}\right) R_{i-1}^{-(2l+1)} \\ 
l\left( \varepsilon _{i}-\varepsilon _{i-1}\right) R_{i-1}^{2l+1} & 
\varepsilon _{i}l+\varepsilon _{i-1}\left( l+1\right)
\end{array}
\right] .
\end{equation}
Similarly for the $p_{i,l}$ coefficients, $p_{N,l}=0\left( \forall l\right) $
and for $m\leq i<N$%
\begin{equation}
p_{i,l}=\frac{p_{i+1,l}E_{11}^{i,l}+E_{12}^{i,l}}{%
p_{i+1,l}E_{21}^{i,l}+E_{22}^{i,l}},
\end{equation}
where the ${\bf E}-$matrix is 
\begin{equation}
{\bf E}^{i,l}{\bf =}\left[ 
\begin{array}{cc}
\varepsilon _{i}\left( l+1\right) +\varepsilon _{i+1}l & \left( l+1\right)
\left( \varepsilon _{i}-\varepsilon _{i+1}\right) R_{i}^{-(2l+1)} \\ 
l\left( \varepsilon _{i}-\varepsilon _{i+1}\right) R_{i}^{2l+1} & 
\varepsilon _{i+1}\left( l+1\right) +\varepsilon _{i}l
\end{array}
\right] .
\end{equation}
Some interesting features of Eqs. (4)-(11) are: i) the coefficients $q_{i,l}$
and $p_{m,l}$ are all identically zero in the absence of dielectric mismatch
at the dot boundary ($\varepsilon _{in}=\varepsilon _{out}$); ii) the
function $F_{im,l}({\bf r},{\bf r}^{\prime })$ defined in Eqs. (5)-(7) has
four different contributions, corresponding to particle-particle
interactions (the term $r_{<}^{l}$ $/$ $r_{>}^{l+1}$), particle-induced
charge interactions (the two terms either proportional to $q_{i,l}$ or $%
p_{m,l}$), and the induced charge - induced charge interaction (the term
proportional to the product $q_{i,l}$ $p_{m,l}$). Only the first one, which
represents the bare Coulomb interaction, survives in the homogeneous limit $%
\varepsilon _{in}=\varepsilon _{out};$ iii) in the calculation of $%
V_{c}^{i,m}({\bf r},{\bf r}^{\prime }),$ only the $q_{i,l}$'s $(p_{m,l}$'s$)$
with $0<i\leq m$ $(m\leq i<N)$ are necessary. For the exciton problem, $%
e^{\prime }=-$ $e.$

From Eqs. (\ref{ecvc}) and (\ref{ecfmm}), the self-polarization potential
can be calculated by taking ${\bf r=r}^{\prime },$ $e=e^{\prime },$
eliminating from (\ref{ecfmm}) the direct contribution $r_{<}^{l}$ $/$ $%
r_{>}^{l+1},$ and dividing by 2 as corresponds to a self-energy. The result
is 
\begin{eqnarray}
V_{s}^{m}\left( {\bf r}\right) &=&\frac{e^{2}}{2\varepsilon _{m}}\text{ }%
\sum\limits_{l=0}^{\infty }\frac{1}{\left( 1-p_{m,l}q_{m,l}\right) }\times 
\nonumber \\
&&\left[ p_{m,l}r^{2l}+p_{m,l}q_{m,l}r^{-1}+q_{m,l}r^{-2\left( l+1\right)
}\right] .  \label{ecvs}
\end{eqnarray}
With $m$ running from $0$ to $N,$ the self-polarization potential $V_{s}(%
{\bf r})$ is defined in the whole space.

It is straightforward to check that in the limit of $\delta =0$ $\left(
N=1\right) $ Eqs. (\ref{ecvc}) and (\ref{ecvs}) reduced to the usual
expressions for $V_{c}$ and $V_{s}$ used in the study of spherical QD's with
boundary dielectric mismatch.\cite{brus,lannoo1,chino}

The choice of different profiles for the dielectric interface modify both
the strength and the functional form of the potentials. In Fig. 2 we show
the self-polarization potential (Fig. 2a) and the $l=0$ component of the
generalized Coulomb potential\cite{selection} (Fig. 2b) for a small quantum
dot where $R=10$ \AA , $\varepsilon _{in}=12.6$ (GaAs), $\varepsilon
_{out}=1 $ (vacuum). The dot-dashed curves correspond to the case of the
step-like model for $\varepsilon \left( r\right) $ (that is $\delta =0$ and $%
N=1$). The self-polarization potential for this case exhibits the already
discussed divergent behavior at the interface, which leads to a divergent
self-energy for the finite barrier situation. The dotted curve is the result
of choosing a linear model for the dielectric interface between $\varepsilon
_{in}$ and $\varepsilon _{out}$, and the solid line corresponds to a profile
proportional to a cosine function; in both cases, $\delta =5$ \AA\ and $%
N=500 $.

\begin{figure}[tbp]
\centering
\epsfxsize=8.75cm
\leavevmode
\epsffile{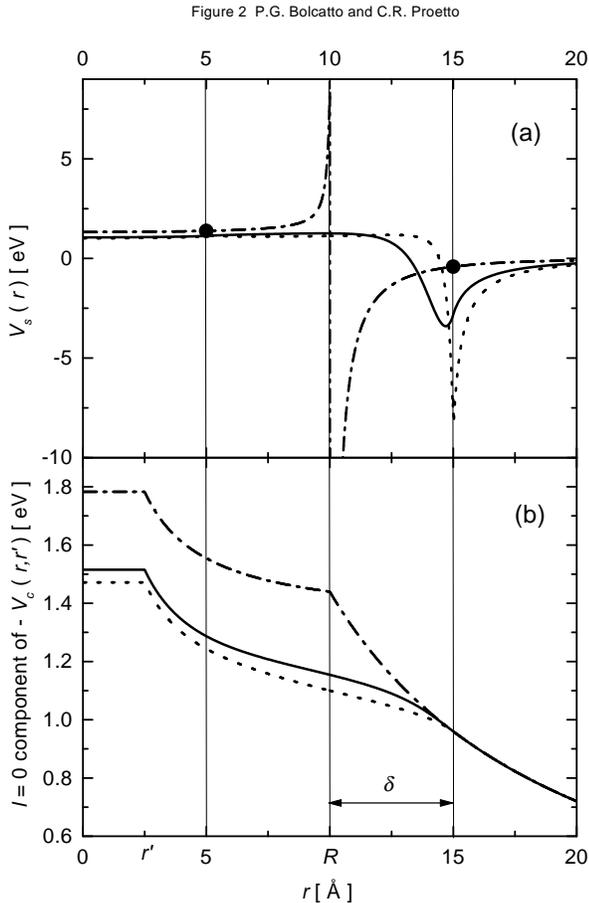}
\vskip1mm
\caption{Self-polarization potential (upper pannel), and $l=0$ component of
the generalized Coulomb potential (lower pannel) as a function of $r,$ for a
quantum dot with $R=10$ \AA , and several models of the dielectric
interface. Dot-dashed line, $\delta =0$; dotted line, $\delta =5$ \AA\
(linear profile); solid line, $\delta =5$ \AA\ (cosine-like profile). For
the generalized Coulomb potential, the source charge is located at $%
r^{\prime }=2.5$ \AA .}
\end{figure}

We can observe in Fig. 2a that the consideration of a finite
interface [independently of the form of $\varepsilon (r)$] eliminates the
divergence present when $\delta =0$. It is worth to emphasize here the
important difference between the ``regularized'' self-polarization potential
and the solutions which arise from the finite size dielectric interface.
Faced with the dilemma of a divergent self-polarization potential and
accordingly a divergent self-energy, no physical guidance (except that the
final result should be finite) helps one in choosing a regularization
procedure for the step-like model of the dielectric interface. In our
approach, we do have such a physical guidance: the dielectric function must
change smoothly in moving from the dot towards the matrix (presumably on the
scale of a few \AA ), and from this physically plausible assumption the
non-divergent self-polarization potential emerge naturally from calculation.
The lack of a physical guidance for the case of a step-like model of the
dielectric interface is clearly reflected in the fact that the
``regularized'' self-polarization potentials of previous works\cite
{kumagai,tran,ban} are quite different from our physically correct
self-polarization potentials shown in Fig. 2. For instance, a regularized
potential following the criteria of Ref. [\onlinecite{ban}] should be made by 
replacing the dot-dashed curve for 5 \AA $<R<$ 15 \AA  by a straight line
connecting the black dots.

Similarly to the planar semiconductor-insulator interface, the linear model
for the dielectric interface leads to a potential which exhibits singular
points at the left and right edge of the interface,\cite{kink} precisely
where the derivative of $\varepsilon (r)$ is not continuous. While from our
numerical approach it is difficult to determine the type of singularity, it
can be integrated easily numerically, so we can safely state that the
singularity is integrable. However, a cosine-like function seems to be a
more physical choice because it gives rise to a well-defined
self-polarization potential everywhere. An important feature of this
potential for all cases is that it can be either repulsive or attractive.
When the source charge is located in the region with greater dielectric
constant, the induced charge has the same sign as the source and then the
interaction is repulsive. On the contrary, if the source charge is placed in
the zone with lower dielectric constant, the induced charge has opposite
sign and the interaction is attractive. Note however the strong asymmetry
between the repulsive and attractive potential strengths. In next section,
we will see that this singular behavior has a relevant role on the
calculation of the excitonic energies.

Regarding the generalized Coulomb potential $V_{c},$ all the curves have the
universal expected $r^{-1}$ decay screened by $\varepsilon _{out}$ outside
the dot $\left( r>R+\delta \right) .$ The results for linear- or cosine-like
profiles for $\varepsilon \left( r\right) $ are quite similar (dotted and
solid lines, respectively), being the potential strength in both cases
smaller as compared with the step-like model. This can be explained as
follows; from Eq. (4), the $l=0$ component of the generalized Coulomb
potential for the step-like model for the dielectric interface is 
\begin{equation}
V_{c}^{0,0}({\bf r},{\bf r}^{\prime })=-\text{ }\frac{e^{2}}{\varepsilon
_{in}R}\left( \frac{R}{r^{\prime }}+\frac{\varepsilon _{in}-\varepsilon
_{out}}{\varepsilon _{out}}\right)
\end{equation}
for $r<r^{\prime }$ (the region where the potential is constant of Fig. 2b).
The first term in Eq. (13) corresponds to the bare Coulomb potential, the
second to the potential generated by the charges induced at the dot
boundary. Note that this term is proportional to the difference of
dielectric constants at both sides of the interface; consequently, by giving
a finite size to the dielectric interface this term can only decrease in
magnitude, as the difference between two successive dielectric constants
approach zero for $N\rightarrow \infty $. Already from the results presented
in Fig. 2b a decrease of the Coulomb energy of the exciton for the case of a
finite size dielectric interface can be anticipated, as compared with the $%
\delta =0$ interface.

\subsection{Zero-order eigenvalues and eigenfunctions}

As we will consider quantum dots with characteristic dimensions smaller than
the exciton effective Bohr radius, we will use the strong confinement
approximation (SCA) for the calculation of the eigenvalues, wave functions,
and excitonic energies.\cite{efros,brus} In this approach the
self-polarization and Coulomb interactions (which scales as $R^{-1}$) are
considered as a perturbation against the kinetic energy (which scales as $%
R^{-2}$). According to this, we propose a separable wave function for the
electron-hole system $\Psi \left( {\bf r}_{e},{\bf r}_{h}\right) \cong \psi
_{e}\left( {\bf r}_{e}\right) \psi _{h}\left( {\bf r}_{h}\right)
=R_{nl}\left( r_{e}\right) Y_{lm}\left( \theta _{e},\varphi _{e}\right)
R_{n^{\prime }l^{\prime }}\left( r_{h}\right) Y_{l^{\prime }m^{\prime
}}\left( \theta _{h},\varphi _{h}\right) $, where $R_{nl}(r_{i})$ are the
radial component of the wave-function, and $Y_{lm}(\theta ,\varphi )$ are
spherical harmonics. As a consequence, the resolution of the zero-order
Hamiltonian consists of two equivalent problems 
\begin{eqnarray}
H_{i}^{\left( 0\right) }\left( {\bf r}_{i}\right) \psi _{i}\left( {\bf r}%
_{i}\right) &=&\left[ -\frac{\hbar ^{2}}{2m_{i}\left( {\bf r}_{i}\right) }%
\nabla _{i}^{2}+V_{0i}\left( {\bf r}_{i}\right) \right] \psi _{i}\left( {\bf %
r}_{i}\right)  \nonumber \\
&=&E_{i}\psi _{i}\left( {\bf r}_{i}\right) ,  \label{eczero}
\end{eqnarray}
where $i=e,h.$

The complete solutions of Eq. (14) turned out to be spherical Bessel
functions of real and imaginary arguments;\cite{schiff} focusing the
attention on the ground state $\left( n=1,\text{ }l=m=0\right) $, the
angular contribution of $\psi _{i}\left( {\bf r}_{i}\right) $ becomes a
constant and the normalized solution of Eq. (\ref{eczero}) is 
\begin{eqnarray}
\psi _{i}\left( {\bf r}_{i}\right) &=&\frac{A_{i}}{\sqrt{4\pi }}\left[
\Theta \left( R-r_{i}\right) \frac{\sin \left( k_{i}^{in}r_{i}\right) }{r_{i}%
}+\right.  \nonumber \\
&&\left. \Theta \left( r_{i}-R\right) B_{i}\frac{\exp \left(
-k_{i}^{out}r_{i}\right) }{r_{i}}\right] ,  \label{ecpsi}
\end{eqnarray}
where the eigenvalues are defined through $k_{i}^{in}=(2m_{i}^{in}E_{i}/%
\hbar ^{2})^{1/2}$ and $k_{i}^{out}=\left[ 2m_{i}^{out}\left(
V_{0i}-E_{i}\right) /\hbar ^{2}\right] ^{1/2}.$ As the electron (or hole)
has different effective mass depending on the material, the wave function
must verify the Ben Daniel-Duke boundary conditions,\cite{bastard} i.e. $%
\psi _{i}\left( R^{-}\right) =\psi _{i}\left( R^{+}\right) $ and $\psi
_{i}^{\prime }\left( R^{-}\right) /m_{i}^{in}=\psi _{i}^{\prime }\left(
R^{+}\right) /m_{i}^{out}.$ By applying this special boundary conditions we
arrive to the following implicit eigenvalue equation for the quantum dot
energies 
\begin{equation}
k_{i}^{in}R\cot \left( k_{i}^{in}R\right) =1-\frac{m_{i}^{in}}{m_{i}^{out}}%
\left( 1+k_{i}^{out}R\right) .
\end{equation}
$A_{i}$ and $B_{i}$ are constants determined by normalization requirements
and are equal to $A_{i}=\left[ R/2-\sin \left( 2k_{i}^{in}R\right)
/4k_{i}^{in}+\sin ^{2}\left( k_{i}^{in}R\right) /2k_{i}^{out}\right] ^{-1/2}$
and $B_{i}=\exp \left( k_{i}^{out}R\right) \sin \left( k_{i}^{in}R\right) .$
An important point in this treatment is that we maintain a sharp-profile
(step-like model) for the electronic interface, choosing its boundary at the
midpoint of the dielectric interface; a more rigorous treatment should also
give a finite size to the electronic interface. Our approximation is
justified for not too small QD's, as then the electronic interface is a
small fraction of the total dot size, and accordingly zero-order eigenvalues
and wave-functions are weakly dependent on the finite size of the interface
(typically a few angstroms).

\begin{figure}[tbp]
\centering
\epsfxsize=8.75cm
\leavevmode
\epsffile{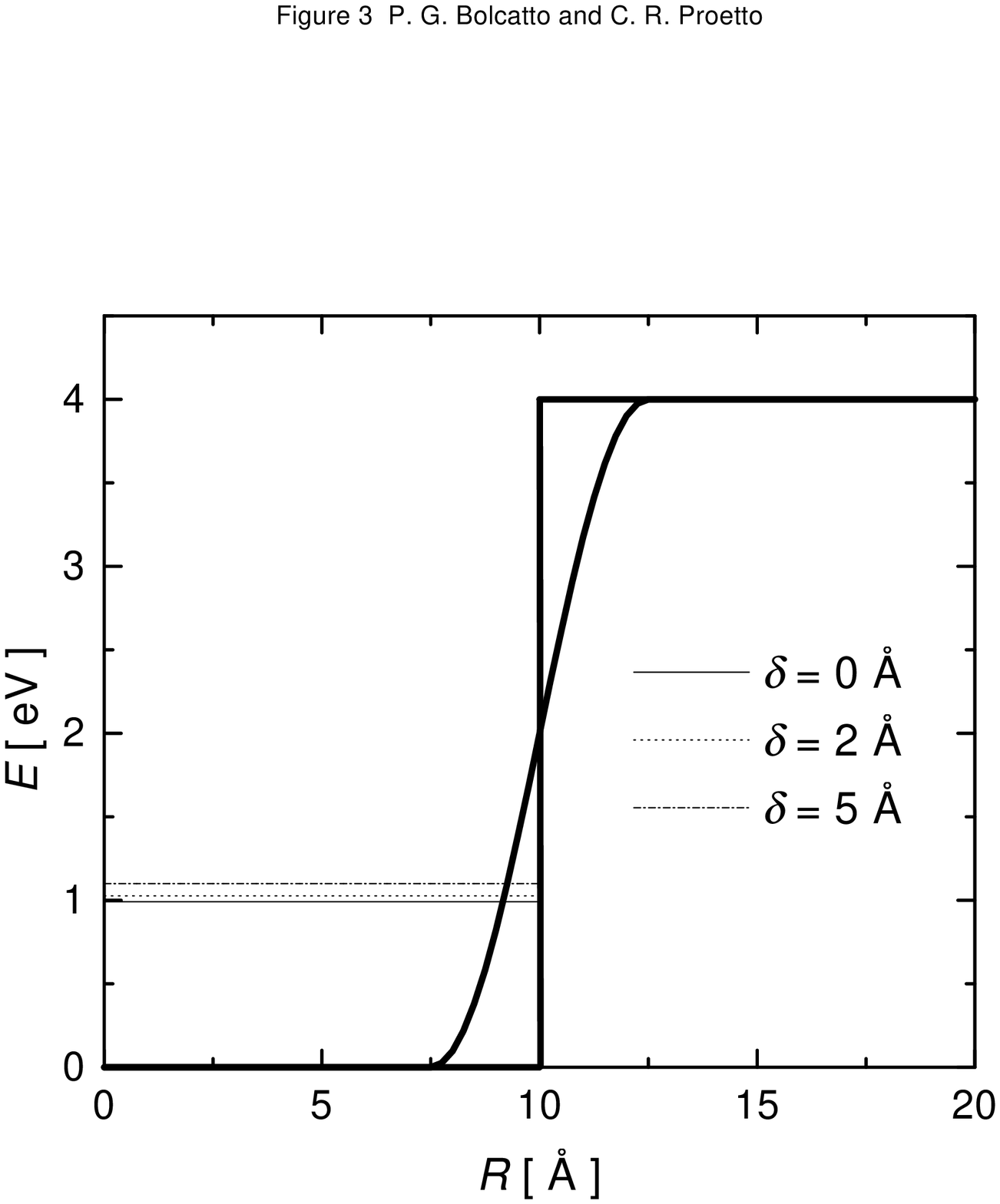}
\vskip1mm
\caption{Bound-state eigenvalues (thin lines) of the quantum dot spherical
well, for the step- and cosine-like models (thick lines) of the electronic
interface. For $\delta =0$ $\left( \delta \neq 0\right) $ the eigenvalues
are exact (approximated), as explained in the text. For this particular dot
size, only one bound state exist in the spherical well.}
\end{figure}

In order to give a quantitative support to this approximation for the
electronic interface, we present in Fig. 3 a calculation of the first-order
correction to the square-well eigenvalues, when the confining potential is
modified in correspondence with the finite size of the dielectric interface.
In particular, we adopt for the finite-size electronic interface the same
cosine-like profile as for the finite-size dielectric interface.
Accordingly, full, dashed-dotted, and dotted lines correspond to the
cosine-like electronic interface with $\delta =0,$ $2,$ and $5$ \AA . In the
last two cases, the eigenvalues were obtained treating as a perturbation the
difference between the cosine- and step-like electronic profiles. In the
worse case of $\delta =5$ \AA , the correction amounts to about 10 \%, while
for $\delta =2$ \AA\ (not shown) is much smaller ($\sim $ 3 \%). Similar
differences could be expected for the perturbed wave functions. Is
interesting to note that as the perturbed eigenvalue is above than the
corresponding to the step-like electronic profile, one can also expect a
greater penetration in the barrier. However, the decay of the perturbed wave
function starts before than the decay of the unperturbed one, and
consequently some compensation exist between both effects. Finally, the
example displayed in Fig. 3, with $\delta /R=0.5$ (for $\delta =5$ \AA ) can
be considered as the less favorable configuration for this approximation, as
the error decreases with $\delta /R.$ For instance, for $\delta /R=5$ \AA\ / 
$20$ \AA\ $=0.25,$ the difference between perturbed and unperturbed
eigenvalues is only about 5 \%.

\subsection{Exciton Coulomb energy}

The exciton Coulomb energy is defined as the correction to the
single-particle size-dependent band gap energy needed to create an
electron-hole pair inside the quantum dot. Then, this excitonic energy
includes the generalized Coulomb interaction between the negative and
positive charges as well as the self-polarization energies corresponding to
the electron and the hole,

\begin{equation}
E_{ex}\equiv \text{ }\Sigma _{e}+\Sigma _{h}-E_{Coul}.  \label{ecexciton}
\end{equation}
The Coulomb energy $E_{Coul}$ and the self-polarization ones $\Sigma _{e}$
and $\Sigma _{h}$ are calculated within the SCA by performing the average
value of $V_{c}$ and $V_{s}$ [Eqs. (\ref{ecvc}) and (\ref{ecvs}),
respectively] using the zero-order wave functions given by (\ref{ecpsi})\cite
{shape}

\begin{equation}
\Sigma _{i}\equiv \int d{\bf r}_{i}\text{ }\psi _{i}^{*}\left( {\bf r}%
_{i}\right) \text{ }V_{s}\left( {\bf r}_{i}\right) \text{ }\psi _{i}\left( 
{\bf r}_{i}\right) ,  \label{ecself}
\end{equation}
\begin{equation}
E_{Coul}\equiv -\int d{\bf r}_{e}d{\bf r}_{h}\left| \psi _{e}\left( {\bf r}%
_{e}\right) \right| ^{2}\text{ }V_{c}\left( {\bf r}_{e},{\bf r}_{h}\right) 
\text{ }\left| \psi _{h}\left( {\bf r}_{h}\right) \right| ^{2}.
\label{eccoul}
\end{equation}
Defined in this way, $E_{Coul}$ for the electron-hole pair is always
positive. For the finite barrier case, $\Sigma _{i}$ can be either positive
or negative, while for the infinite barrier case $\Sigma _{i}$ is always
positive.

\begin{table}[tbp]
\caption{Material parameters used in the calculation.}$
\begin{array}{cccccc}
& m/m_{0} &  & \varepsilon &  & V_{0}\text{ (eV)} \\ 
& \text{in} & \text{out} & \text{in} & \text{out} &  \\ 
\text{electron} & 0.067 & 1 & 12.6 & 1 & 4 \\ 
\text{hole} & 0.13 & 1 & 12.6 & 1 & \infty
\end{array}
$%
\end{table}

\section{Results}

The results of this section were obtained using the material parameters
summarized in Table I, corresponding to a GaAs QD immersed in vacuum. We
choose this system by several reasons: i) the choice of vacuum as the
surrounding media simplifies considerably the material parameter choice, as
then $V_{0e}$ is given by the electron affinity of the semiconductor dot
material in bulk, while for our simple two-band semiconductor model, $%
V_{0h}\rightarrow \infty ;$ ii) the high dielectric mismatch between GaAs
and vacuum maximize the effect of the induced charge at the dot boundary in
which we are interested. An important point to remark is that all the
constants characterizing the materials were extracted from the bulk values
and consequently they are not adjustable parameters. Also, all the results
to be presented below for the self-polarization, Coulomb, and excitonic
energies were obtained by using the cosine-like model for the dielectric
interface; quite similar results were obtained for the linear model of the
dielectric interface.

\begin{figure}[tbp]
\centering
\epsfxsize=8.75cm
\leavevmode
\epsffile{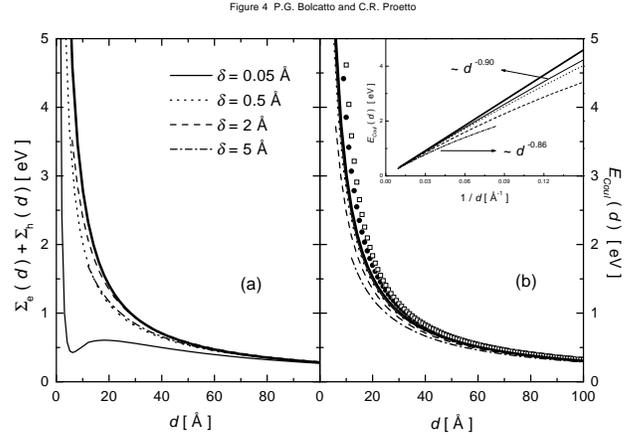}
\vskip1mm
\caption{Exciton self-polarization energies (left pannel) and Coulomb
energies (right pannel) as a function of dot size, for $\delta =$ 0.05, 0.5,
2, and 5 \AA . The inset of Fig. 4b corresponds to $E_{Coul}(d)$ as a
function of $d^{-1}.$ The full thick line corresponds to the infinite
barrier limit for electrons (and $\delta =0$), and the remaining curves are
for $V_{0e}=4$ eV. Full circles and open squares correspond to the
ground-state binding energy of an on-center donor impurity.}
\end{figure}

The lower panel of Fig. 4 shows the results for the Coulomb energy [Eq. (\ref
{eccoul})], for several sizes of the dielectric interface, as a function of
the dot diameter $d=2R.$ As a reference we include a thick curve
corresponding to the usually quoted result\cite{brus} $E_{Coul}(d)\simeq
3.786$ $e^{2}/d\varepsilon _{in}$ which results from the combination of
infinite barriers $\left( V_{0e}\rightarrow \infty \right) $ with step-like
model for the dielectric interface. We can observe that, as was expected
from the behavior of $l=0$ component of $V_{c}$ shown in Fig. 2b, this
contribution to the exciton energy decreases monotonously when the thickness
of the dielectric interface increases. For a better visualization of the
scaling with dot size we plot in the inset the variation of $E_{Coul}$ with
the reciprocal diameter $1/d$. The thin line corresponds to $\delta =0.05$
\AA\ which can be taken as a very good approximation to the $\delta =0$
case. As it was established in Ref. [\onlinecite{fema}], as soon as we go
apart from the infinite barrier approximation, the widely criticized $d^{-1}$
scaling of IEMA is lost. Here, we have demonstrated that it is possible to
modify even more the size dependence if we consider a finite size dielectric
interface connecting the QD material and the surroundings. For instance, for 
$\delta =5$ \AA\ the scaling of $E_{Coul}$ with dot size is proportional to $%
d^{-0.86}$.

The exciton self-polarization energy given by $\Sigma _{e}+\Sigma _{h}$
plotted on the upper panel of Fig. 4 shows an interesting behavior with a
non-monotonic dependence on the dielectric interface size. As in the lower
panel, the full thick line corresponds to the infinite barrier limit
combined with the step-like model of the dielectric interface; accordingly,
the difference between the full thick line $(\delta =0,$ $V_{0e}\rightarrow
\infty )$ and the full thin line $(\delta =0.05$ \AA $,V_{0e}=4$ eV$)$ is
due to the effect of the finite barrier size for the latter. In particular,
polarization effects are strongly suppressed for near zero-width interfaces,
afterwards these effects are augmented up to $\delta =2$ \AA ,\ but for
wider interfaces the self-polarization energy decays again. This singular
feature can be understood if we remember that the self-polarization energies
actually are the product (integrated over the whole space) between the
self-polarization potential $V_{s}$ and the electron (hole) probability
density given by $\left| \psi _{e}\right| ^{2}\left( \left| \psi _{h}\right|
^{2}\right) $ [Eq. (\ref{ecself})].

To help with the explanation we present
the Fig. 5 where $\left| \psi _{e}\right| ^{2}$ and $V_{s}$ are plotted
separately. As can be immediately grasped from this figure, the maximum
value for the self-polarization energy correction for the combination of
total confinement and step-like dielectric interface is a consequence of the
fact that the wave-function is different from zero precisely in the same
region where the self-polarization potential is positive (inside the dot).
The situation changes drastically by relaxing the hard-wall boundary
condition, as then $\left| \psi _{e}\right| ^{2}$ spreads considerably
outside the dot, where the self-polarization potential is negative. This
explains the strong decrease of $\Sigma _{e}+\Sigma _{h}$ by going from $%
V_{0e}\rightarrow \infty $ to $V_{0e}=4$ eV$;$ actually, as $\Sigma _{h}$ is
always positive (the holes are perfectly confined in the present case) an
almost zero value of $\Sigma _{e}+\Sigma _{h}$ implies a negative value for $%
\Sigma _{e}.$ As the dielectric interface increases in size, the negative
part of the self-polarization potential moves apart from the electronic
interface, $\Sigma _{e}$ becomes less negative, and $\Sigma _{e}+\Sigma _{h}$
increases. Finally, for $\delta \gtrsim 3$ \AA , only the positive component
of $V_{s}$ contributes to $\Sigma _{e},$ and because its value decreases as $%
\delta $ increases, $\Sigma _{e}$ shows a non-monotonic behavior. A final
remark about Fig. 4 is that both $E_{Coul}$ and $\Sigma _{e}+\Sigma _{h}$
values never exceed those achieved with the total confinement approximation,
which seems to be a kind of upper limit.

\begin{figure}[tbp]
\centering
\epsfxsize=7.75cm
\leavevmode
\epsffile{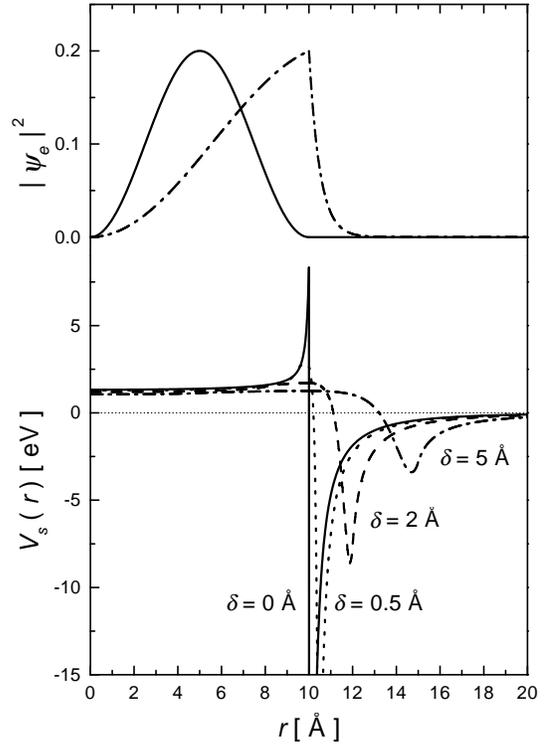}
\vskip1mm
\caption{Upper pannel: probability density associated with the normalized
ground state electronic wave-function, and effective mass mismatch $%
m_{in}/m_{out}=0.067$. Full line, $V_{0e}\rightarrow \infty ;$ dashed-dotted
line, $V_{0e}=4$ eV. Lower pannel: self-polarization potential for $\delta
=0,$ $0.5,$ $2,$ $5$ \AA\ Dot radius, $R=10$ \AA .}
\end{figure}

Concerning the excitonic energy, the results seem to be less predictable. In
Fig. 6 we show the exciton energies $E_{ex}(d)$ as a function of dot sizes,
for quantum dots with $\delta =0.05,$ $0.5$, $2$, and $5$ \AA ; the IEMA
results for $\delta =0$ are also plotted (thick full line). Contrary to $%
E_{Coul}$ and $\Sigma _{e}+\Sigma _{h},$ much higher excitonic energies than
those resulting from IEMA can be reached for QD's with thin interfaces. The
explanation for this is clear from the results presented in Fig. 4: while
the self-polarization energy decreases strongly by going from $%
V_{0e}\rightarrow \infty $ to $V_{0e}=4$ eV$,$ the Coulomb energy remains
essentially unaltered under such change in confinement. Accordingly, the
exciton energy which is the difference between these two magnitudes,
increases and becomes similar in magnitude to the Coulomb contribution
(roughly, $E_{ex}\sim E_{Coul}$ $/$ $2$). As the dielectric interface
increases in size, the self-polarization energy first increases up to a
critical $\delta $, and then decreases: this non-monotonic behavior of $%
\Sigma _{e}+\Sigma _{h}$ is followed by a non-monotonic behavior for the
exciton energy displayed in Fig. 6. It is interesting to note that for
dielectric interface sizes of about a lattice parameter $(2$ \AA\ $\lesssim
\delta \lesssim 5$ \AA $)$, the excitonic energy correction is quite small
for all $d\gtrsim 20$ \AA . Under these conditions, a measure of the optical
band-gap should be compared directly with the single-particle band gap,
without excitonic corrections.

It is worth noting the richness of behaviors than can be obtained for the
dot size dependence of the exciton energy by the combination of partial
confinement with a finite size for the dielectric interface. This should be
kept in mind when correcting single-particle band gaps with the excitonic
contribution, in order of be able of compare with optical experiments, as
this is done usually by using the infinite barrier and step-like model for
the dot-confined exciton (full thick line of Fig. 6).

\begin{figure}[tbp]
\centering
\epsfxsize=8.75cm
\leavevmode
\epsffile{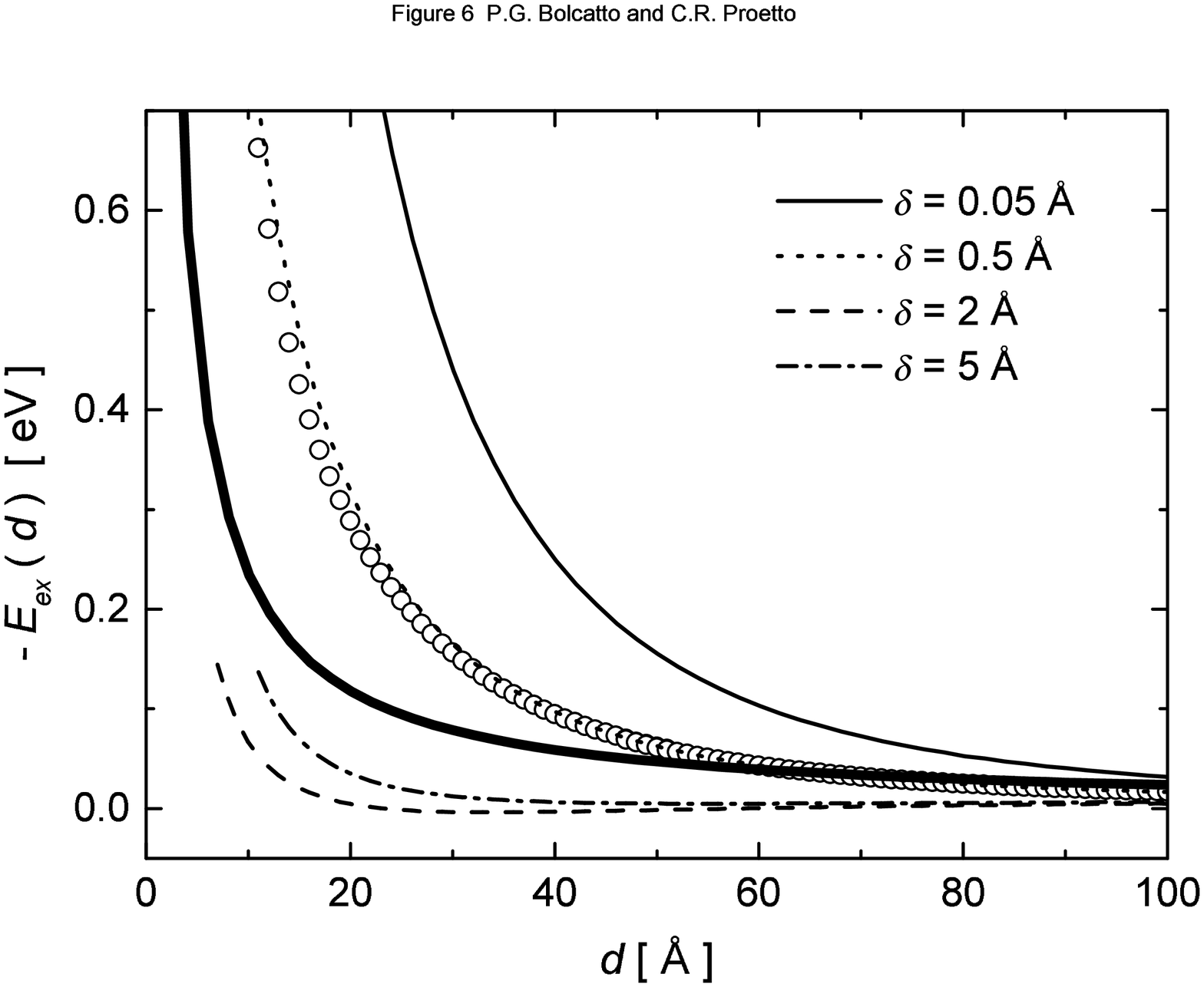}
\vskip1mm
\caption{Exciton energies as a function of dot size for $\delta =$ 0.05,
0.5, 2, and 5 \AA . The full thick line corresponds to the infinite barrier
limit for electrons (and $\delta =0$), and the remaining lines are for a
barrier of 4 eV. As explained in the text, the curve with open points
correspond to the exciton energy obtained by replacing $\varepsilon
_{in}=12.6$ by $\varepsilon _{in}(d)$ as defined by Eq. (20).}
\end{figure}

Before conclude, let us take again the issue of the electrostatic screening
in semiconductor quantum dots, discussing it on the light of the preceding
results. In the first place, we can include easily in our calculations a
size-dependent dielectric constant for the dot. According to the generalized
Penn model,\cite{tsu} 
\begin{equation}
\varepsilon _{in}(d)=1+\frac{\varepsilon _{in}-1}{1+(2\alpha \text{ }/\text{ 
}d)^{l}}
\end{equation}
with $\varepsilon _{in}$ being the infinite dot size (bulk) dielectric
constant, $\alpha =2\pi E_{F}/E_{g}k_{F},$ and $l=2.$ $E_{g}$ is taken as
the optical gap of the dot-acting semiconductor, and $k_{F}$ and $E_{F}$ are
determined from the density of valence electrons $n_{0}.$ For GaAs, taking $%
E_{g}=5.2$ eV,\cite{yu} and $n_{0}=0.1774/$\AA $^{3}$, we obtain $\alpha
=8.01$ \AA . Inserting this size-dependent dielectric constant in our
calculations, we obtain the curve with open points of Fig. 6 (corresponding
to $\delta =0.5$ \AA ). It can be seen that this size effect, {\it although
it exist}, is quite small for dots with $d\gtrsim 10$ \AA .

The related issue of the incomplete screening at short distances in small
semiconductor clusters can be discussed along a similar line of reasoning.
Generalizing Eq. (20) as 
\begin{equation}
\frac{1}{\varepsilon _{in}(r,d)}=\frac{1}{\varepsilon _{in}(d)}+\left( 1-%
\frac{1}{\varepsilon _{in}(d)}\right) e^{-\text{ }r/a},
\end{equation}
where $\varepsilon _{in}(d)$ is given by Eq. (20), and $a$ is a screening
parameter. For $d\rightarrow \infty $, and $\varepsilon _{in}(\infty )$
corresponding then to the bulk dielectric constant, the spatially dependent
dielectric function of Eq. (21) is that proposed by Hermanson.\cite
{hermanson} The screening parameter $a$ can be obtained from the criteria
that the Fourier transform of Eq. (21) fits the dielectric function of Walter
and Cohen.\cite{walter} Following Oliveira and Falicov,\cite{oliveira} we
took $a\simeq 0.58$ \AA\ as the characteristic value for GaAs. Without any
calculation, we can argue against the importance of this effect due to the
following facts: i) the spatially dependent screening, as defined by Eq.
(21) for a given dot size is extremely short-ranged, with a typical length
scale of about an {\it atomic} Bohr radius, and ii) according to Eq. (20), $%
\varepsilon _{in}(d\rightarrow 0)\rightarrow 1,$ and then {\it the
correction given by the second term in Eq. (21) is smaller for smaller dots}%
. In other words, in this limit $\varepsilon _{in}(r,d\rightarrow
0)\rightarrow \varepsilon _{in}(d\rightarrow 0),$ and the incomplete
short-range screening reduces to the size-dependent dielectric function. Let
us estimate quantitatively the importance of the second term in Eq. (21) on
the ground state binding energy of unconfined (bulk) excitons: using
first-order perturbation theory the correction (in units of the bulk exciton
binding energy $R_{0}^{*}$) is given by 
\begin{equation}
\frac{\Delta E}{R_{0}^{*}}=\frac{8\left[ \varepsilon _{in}-1\right] }{\left(
2+a_{0}^{*}/a\right) ^{2}},
\end{equation}
with $R_{0}^{*}=e^{2}/2\varepsilon _{in}a_{0}^{*},$ $a_{0}^{*}=\hbar
^{2}\varepsilon _{in}/e^{2}\mu ,$ and $\mu =m_{e}m_{h}/(m_{e}+m_{h}).$
Inserting in Eq.(22) the GaAs parameters of Table I, we obtain $\Delta
E/R_{0}^{*}\simeq 10^{-3}.$ Having discarded the importance of incomplete
short-range screening for the bulk, let us estimate its relevance for
quantum dots. To this end, we have calculated this correction for the case
of the ground state binding energy of an on-center donor impurity. This
could be considered as the limiting case of $m_{h}/m_{0}\gg 1,$ as in this
case the exciton and the donor impurity problems are equivalent. In order to
simplify matters, we take here the infinite barrier approximation. The
results are given by the upper two curves in Fig. 4b. Of these two, the
lower one (circles) corresponds to use in the calculation of $E_{Coul}(d)$ 
for the donor impurity only the first term in the right-hand side of Eq.(21), 
while the upper curve (squares) is the result by using Eq.(21) as it is. 
Once again,
short-range effects are found quite small for dot sizes $d\gtrsim 20$ \AA .

\section{Conclusions}

In this work we have studied the dependence of Coulomb, self-polarization,
and excitonic energies of spherical semiconductor quantum dots with partial
confinement on dot size and in the presence of a smooth dielectric interface
at the dot boundary. We have developed a numerical method which allows the
study of the electrostatic properties of finite-size dielectric interfaces
of arbitrary shape; essentially, it consist in replacing the continuous
profile of the dielectric interface by a series of discrete steps. In the
limit of very large number of steps (several hundreds), our method provides
the exact solution to the electrostatic problem of a graded
three-dimensional dielectric interface. The motivation for this analysis is
twofold: from the mathematical point of view, the combination finite barrier
plus step-like model for the dielectric interface is ill-behaved, as the
self-energy of the partially confined particle (electron, hole, or both)
diverges in this situation. From the physical point of view, the interface
between two semiconductors (or between semiconductor and vacuum) is never
perfectly sharp as the step-like model of the dielectric interface assumes.
The finite size of the interface becomes then a crucial feature for a
sensible calculation of the self-energy, although is not so relevant for the
Coulomb energy.

We have studied the excitonic energy for linear and cosine-like spatial
profiles for the dielectric interface, and for different values of the
interface width. Interestingly, the excitonic energy shows a pronounced and
non-monotonic dependence on the interface width, becoming quite small for
intermediate dielectric interfaces, but becoming quite sizeable for narrow
dielectric interfaces. Our results highlights the importance of consider the
finite size of the dielectric interface, as quite a variety of excitonic
corrections can be obtained depending on the combination barrier height -
interface width.

\section{Acknowledgments}

One of us (P.G.B.) wishes to acknowledge the financial support received from
FOMEC N$^{\text{o}}$ 331, Universidad Nacional del Litoral. The authors are
grateful Karen Hallberg for a careful reading of the manuscript.


\begin{references}
\bibitem{limite}  For example, for the step-like model of the dielectric
interface, the partial-wave expansion of Eq. (12) diverges as the
geometrical series for $r\rightarrow R,$ but with two different coefficients
depending if $r$ approach $R$ from inside or outside the dot. This means
that the potential singularity at the dot surface is not integrable in the
sense of a principle value integral; more precisely, it leads to a
logarithmic divergence with the sum index $l$ when the sum of Eq. (12) is
attempted for a finite barrier case.
\bibitem{selection}  This is the only component that enters in the
calculation of the Coulomb energy for the ground state exciton; on the
contrary, all the infinite $l$ components should be kept in the calculation
of the self-polarization energy. In practice, when the sum converges, it is
truncated after $l$ reaches a value such that the result remains constant.
\bibitem{kink}  The left kink of the linear model for the dielectric
interface is not discernible in the scale of the drawing of Fig. 2a.
\end{references}
\end{document}